\documentclass[journal=jacsat,manuscript=article]{achemso}
\usepackage[version=3]{mhchem} 
\usepackage{siunitx}
\usepackage{geometry}
\usepackage[utf8]{inputenc}
\DeclareUnicodeCharacter{2212}{-}
\DeclareUnicodeCharacter{03B4}{$\delta$}

\author{Michael Verhage}
\affiliation{Molecular Materials and Nanosystems (M2N) - Department of Applied Physics and Science Education - Eindhoven University of Technology, Eindhoven, Netherlands}
\author{Kees Flipse}
\affiliation{Molecular Materials and Nanosystems (M2N) - Department of Applied Physics and Science Education - Eindhoven University of Technology, Eindhoven, Netherlands} 
\email{*c.f.j.flipse@tue.nl}

\title{Order in disorder: oxygen vacancy driven electronic phase separation of LaNiO$_{3-x}$ epitaxial thin film surface investigated by scanning probe microscopy}

\begin{document}

\begin{abstract}
Oxygen vacancies in nickelates are known to introduce a variety of emergent phenomena and are considered to significantly affect conductivity. Few studies have examined real-space evidence for oxygen vacancies on the surface, particularly using scanning probe microscopy. Understanding the surface composition, both structurally and chemically, is crucial for the application of nickelates, such as in electrochemical water splitting reactions as catalysts. In this study, we investigate a \SI{20}{\nano\meter} epitaxial LaNiO$_{3-x}$ (LNO) film grown on SrTiO$_3$ (STO) via pulsed laser deposition. We examine the surface of this film using scanning tunneling microscopy (STM) and reveal a complex surface morphology composed of densely packed crystalline nanometer-sized circular features with radius ranging from \SI{10}{} to \SI{20}{\nano\meter}. Scanning tunneling spectroscopy revealed an electronic inhomogeneity, or phase separation, in nanoscale islands of semi-conductive nature embedded within a metallic matrix. This change in the electronic density of the states was associated with increased concentration of oxygen vacancies. Further evidence for significant oxygen vacancies at the surface was inferred by examining scanning tunneling tip-induced surface degradation. The strong electric field between the tip and sample likely facilitates oxygen removal in a vacuum environment, increasing the formation of vacancies, which can lead to the breakdown of the crystal structure. This provides insight into the possible origin of the chemical surface transformation during the electrochemical water splitting reaction of this nickelate.
\end{abstract}


\section{Introduction}

LaNiO$_3$ (LNO) is a distinctive complex oxide perovskite that is part of the $Re$NiO$_3$ family, where $Re$ denotes a rare-earth element. Studies on the electronic and chemical inhomogeneities of LNO thin films, especially at the surface, have been relatively sparse compared to other rare-earth nickelates. This may be attributed to the widespread view of LNO as a Pauli paramagnet \cite{Torrance1992SystematicGap}, which does not undergo a metal-to-insulator transition and remains metallic even at very low temperatures. However, recent research has uncovered a complex  behaviour of LNO such as emerging magnetic ordering \cite{Asaba2018UnconventionalLaNiO3}, influenced by factors such as interface engineering \cite{Soltan2023FerromagneticSuperlattices}, strain leading to charge disproportion \cite{Yoo2015LatentTopology, Yoo2016ThicknessdependentStrain}, and film purity driven quantum fluctuations \cite{Liu2020ObservationLaNiO3}. An extensive angle-resolved photoemission spectroscopy (ARPES) study by Yoo \textit{et al.} \cite{Yoo2015LatentTopology} has shown a clear link between strain and the onset of charge- and bond disproportionation. A phase separation into metallic and insulating phases is known in different RNiO$_3$, but only in the vicinity of the MIT \cite{Mattoni2016StripedNickelates, Preziosi2018Direct3}. For thin films of LaNiO3/LaAlO$_3$ Karners spin lattice relaxation measurements indicate microscopically separated regions with distinct electronic properties \cite{Karner2021EvolutionNMR}. 

The study of electronic anomalies, particularly on the surfaces of epitaxial thin film LNO, is driven by its potential as a catalyst for electrochemical reactions like the oxygen evolution reaction (OER) \cite{Asaba2018UnconventionalLaNiO3, Anjeline2022ProbingProperties, Zhao2020BoostingDecoration}. Specifically, oxygen vacancies have been shown to enhance the OER activity of nickelates \cite{Anjeline2022ProbingProperties}, although oxygen vacancy-rich LNO increases the electrical resistivity by orders of magnitude \cite{Kotiuga2019CarrierVacancies}. To gain a deeper insight into the electrochemical mechanisms of these nickelates, it is essential to understand the surface structure and nanoscale electronic inhomogeneity at room temperature. Surface transformations \cite{Baeumer2021TuningElectrolysis} have been found to have a significant impact on OER, with oxygen vacancies potentially playing a crucial role in the degradation and transformation of the pre-catalyst.

In this study, we employ scanning probe microscopy to examine the surfaces of nickelate epitaxial thin films produced by pulsed laser deposition (PLD) \cite{Baeumer2021TuningElectrolysis}, uncovering notable nanoscale corrugations. This finding starkly contrasts with the widely held belief of surfaces being long-range atomically smooth with A-plane or B-plane terminations. Significant nanoscale corrugation may also apply to other metal oxide perovskite epitaxial thin films fabricated by PLD; such nanoscale corrugated surface morphology likely affects the stoichiometry, especially concerning oxygen vacancies.

In this study, we use correlated SPM methods by atomic force microscopy (AFM), scanning tunneling microscopy (STM), and scanning tunneling spectroscopy (STS). These nanoscale imaging techniques revealed that the surface of our LNO film comprises of stacked circular feature-like features with a diameter ranging from 2 to \SI{15}{\nano\meter}. Atomic-scale resolution was attained using STM showing the circular features exhibit a crystalline structure. These insights challenge the perception of long-range atomic flatness of the films. 

Secondly, the local density of states (LDOS) at the surface was mapped to explore potential oxygen vacancies and surface non-stoichiometry. The resistivity of nickelate films has been reported to increase significantly with the concentration of oxygen vacancies \cite{Kotiuga2019CarrierVacancies}. Using STS to map the LDOS of our LNO thin film, nanoscale electronic phase separation was observed on the surface, likely due to the highly localized ordering \cite{Wang2018AntiferromagneticCrystals} and clustering of oxygen vacancies. The formation of oxygen vacancies changes the nominal valence state of Ni$^{3+}$ to Ni$^{2+}$, leading to the localization of electrons \cite{Pavarini2016Quantum2016}, a phenomenon observed in similar nickelates \cite{Kotiuga2019CarrierVacancies}. The 3$d^8\underline{L}$ configuration of the associated $3+$ valence state, with $L$ the oxygen ligand hole, changes to probably a d$^8$ state with the removal of a oxygen atom, on a disturbed octahedral neighboring site. The charge transfer nature of this nickelate supports the electron-electron correlation \cite{Kotiuga2019CarrierVacancies}, and we conjecture that this is likely related to the flattening of the bands observed with ARPES \cite{Yoo2015LatentTopology} and the increase in the electronic gap by LDOS mapping in our STS experiments. 

Given the probable thermodynamic instability of the oxygen anion in the OER \cite{Binninger2015ThermodynamicCatalysts}, it follows that for LNO, avoiding the formation of oxygen vacancies on the surface could be challenging. Consequently, this could unavoidably result in the degradation of the surface structure under OER conditions, leading to surface transformation in-operando \cite{Baeumer2021TuningElectrolysis}. 

Finally, STM revealed an ongoing change in surface corrosion by continuous scanning as a function of bias voltage. The electric field gradient between the tip and the surface increased as the electric field magnitude decreased. It is conjectured that the removal of oxygen from the surface in a vacuum environment and the formation of vacancies during scanning can cause a collapse of the crystal structure. This effect was found to be present in a thin surface layer of \SI{1}{\nano\meter}, probably due to a gradient of oxygen vacancies from the surface to a stoichiometric bulk.   

\begin{figure}[b!]
    \centering
        \includegraphics[scale=0.3]{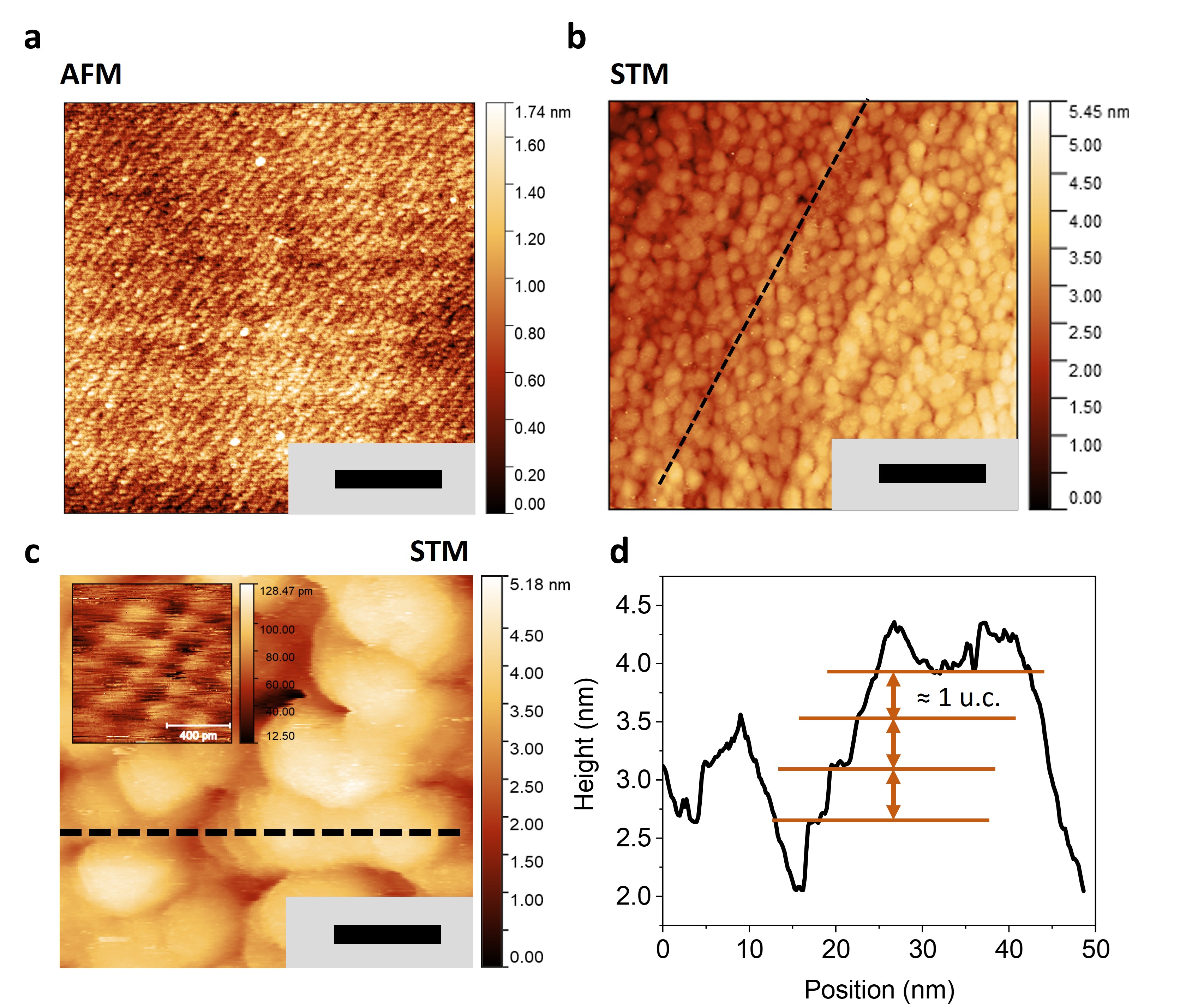}
            \caption{\textbf{Surface characterisation of LNO with AFM and STM.} (\textbf{a}) Topographic AFM image at the micro scale. Nanoscale surface corrugated characteristics are observable with distinct step edges from the vicinal cut STO substrate. The scale bar is equal to \SI{1}{\micro\meter}. (\textbf{b}) STM image, tip bias -\SI{1}{\volt} and tunnel current \SI{10}{\pico\ampere}, revealing a  film surface morphology with circular features. The black dashed lines indicate a step edge. The scale bar is equal to \SI{100}{\nano\meter}. (\textbf{c}) High resolution STM image, tip bias -\SI{150}{\milli\volt} and tunnel current \SI{100}{\pico\ampere}, showing stacked circular features. The black scale bar is equal to \SI{20}{\nano\meter}. The inset shows a small scale atomically resolved STM image, tip bias -\SI{150}{\milli\volt} and tunnel current \SI{100}{\pico\ampere}, of a single circular surface feature, indicating crystalline growth of the film. (\textbf{d}) Topographic line trace of \textbf{c} indicating single unit cell epitaxial growth of the circular features.}
            \label{fig:surface_characterisation}
\end{figure}

These results highlight that unexpected electronic surface phases can exist for LNO thin films and can provide crucial pieces of evidence for deeper understanding on the origin of surface transformation during the electrochemical water splitting of these catalysts \cite{Baeumer2021TuningElectrolysis}, driven by electric field gradients on nanoscale corrugated surfaces leading to oxygen vacancy formation and eventual collapse of the perovskite crystal.  

\section*{Results}
\subsection{Surface characterization with AFM and STM}

LNO films \SI{20}{\nano\meter} thick with dominant La-rich termination were deposited by pulsed laser deposition (PLD) on SrTiO$_3$ (STO) substrates (see Ref. \cite{Baeumer2021TuningElectrolysis} for fabrication details). The STO substrate introduces a tensile strain in the LNO film. We initially chose to utilize micrometer-scale ambient atomic force microscopy (AFM) imaging, as it is frequently employed in the literature to evaluate the quality of similar films. A \SI{5}{}x\SI{5}{\micro\meter} AFM image is presented in \textbf{Fig. \ref{fig:surface_characterisation}a}. Step edges and plateaus, which are characteristic of vicinal cut STO substrates, are visible. However, this image also reveals corrugation, particularly at the nanometer scale. Similar features, with varying degrees of corrugation severity, can be observed in other studies \cite{Baeumer2021TuningElectrolysis, Kante2023HighEntropyOxidation, Fungerlings2023CrystalfacetdependentNickelate}. 

We turned to high-resolution STM imaging, \textbf{Fig. \ref{fig:surface_characterisation}b}, which reveals that the LNO films actually consist of nanoscale circular features, a term we coined before \cite{Verhage2024ElectronicSurface}, with nanoscale holes up to \SI{2.5}{\nano\meter} in depth or around \SI{10}{\percent} of the film thickness. 
The term circular feature used in this work refers to a significant lateral corrugation consisting of spherical features of the film; not an agglomerate of nanoparticles. Note the similarity of the surface corrugation observed with cross-sectional TEM in Refs. \cite{Lopez-Conesa2017EvidenceFilms, Fungerlings2023CrystalfacetdependentNickelate} for similar films. The morphology, distribution, and size of the circular features could depend on the temperature of deposition \cite{Zhu2006FabricationAblation}. 

Furthermore, with AFM we observe that these circular features are evenly distributed on the steps, indicating coherent growth. The edges of the steps can be observed, as indicated by black dotted lines in \textbf{Fig. \ref{fig:surface_characterisation}b}. Our LNO films grown at \SI{650}{\degree}C are La-rich on the surface and not atomically smooth in the sense that only a single-atomic height or a single unit cell (u.c.) height variation is observed across the entire width of the plateaus. The nanoscale imaging indicate significant corrugated nature of the PLD grown LNO films. In previous work, we emphasized the necessity of sharp probes in SPM imaging to correctly assess the smoothness of the film \cite{Verhage2024ElectronicSurface}. The RMS roughness $S_q$ of the films is around \SI{730}{\pico\meter}.

With STM, atomic resolution was observed at several locations on the circular features. For complex transition-metal oxide perovskites, atomically resolved resolution with an STM is limited and has only been reported in a small volume of literature \cite{gai_chemically_2014, Franceschi2020AtomicallyFilms, ronnow_polarons_2006}. Surface corrugations may be difficult to suppress during growth and provide a challenge for atomically resolved STM imaging. Secondly, the resistivity of the metal oxide film, depending on the temperature, may offer limitations to the STM, especially in the insulator phase. In \textbf{Fig. \ref{fig:surface_characterisation}c}-inset an atomically resolved image is given, taken on the surface of a single LNO circular feature. For this image, it is likely that the STM tunnels electrons into the surface oxygen atoms of the LaO surface plane. Furthermore, it was only possible to reveal the atomic surface occasionally and in very small scan areas around \SI{1}{}$\times$\SI{1}{\nano\meter}; probably because the surface curvature requires significant feedback in height of the STM signal, and this can reduce vertical sensitivity. This observation supports the notion that the circular features are crystalline in nature. 

One notable observation from the high-resolution STM images is that the circular features extend into the film's bulk, forming a unified layer, \textbf{Fig. \ref{fig:surface_characterisation}d}. Near the top, significant gaps with depths exceeding \SI{2}{\nano\meter} are visible. It is likely that near the LNO/STO interface, no voids are present, as supported by various TEM studies \cite{Baeumer2021TuningElectrolysis}. This likely indicates that this surface corrugation phenomenon is common in many LNO thin-film studies in the literature.


\subsection{Surface degradation by STM}

\begin{figure*}
    \centering
        \includegraphics[scale=0.3]{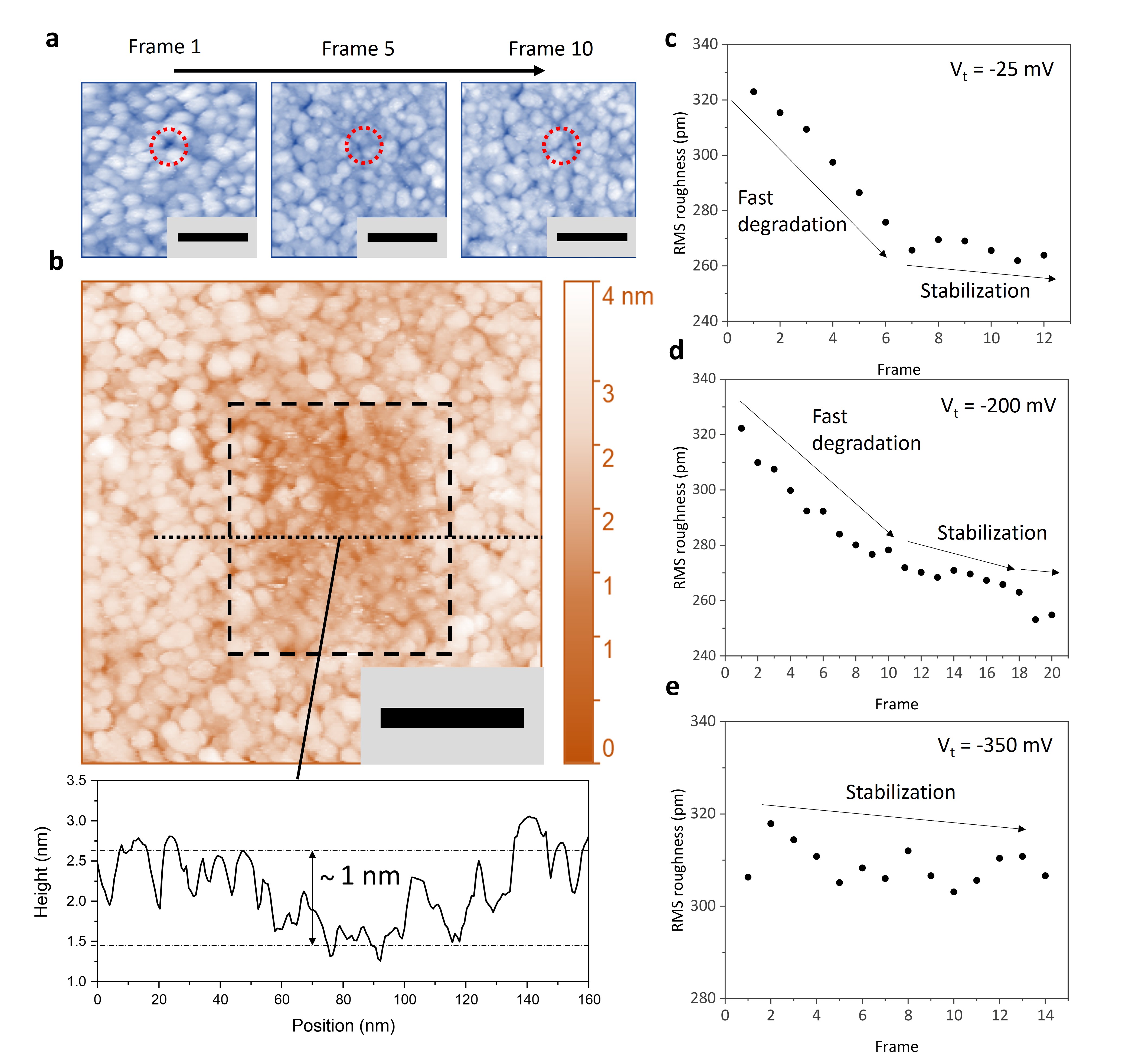}
            \caption{\textbf{Surface corrosion imaged with STM.} (\textbf{a}) Several frames indicating the change in morphology. A central change in feature is highlighted by the red dotted circle. Image taken at tip bias of -\SI{100}{\milli\volt} and tunnel current \SI{100}{\pico\ampere}. The scale bar is equal to \SI{20}{\nano\meter}. (\textbf{b}) Large area scan showing a local square corrosion pattern, introduced by consecutive imaging, tip bias -\SI{25}{\milli\volt}, tunnel current \SI{100}{\pico\ampere}. Approximately \SI{1}{\nano\meter} surface of the LNO film is removed, as indicated with the line trace. The scale bar is equal to \SI{50}{\nano\meter}. (\textbf{c})-(\textbf{e})  Surface roughness (RMS) evolution as function of tip bias voltage in consecutive image frames at the same location. For low tip bias, a fast degradation in surface roughness is followed by stabilization. For higher bias, -\SI{200}{\milli\volt}, the degradation rate is decreased or not observed, -\SI{350}{\milli\volt}.}
            \label{fig:corrosion}
\end{figure*}

A sudden change in surface morphology was detected by continuous STM imaging at the same surface location. The imaging was carried out at room temperature and active compensation for lateral imaging drift was necessary to maintain the tracking of the same morphological features by compensation for drift. In \textbf{Fig. \ref{fig:corrosion}a}, three frames are displayed, selected from \SI{11}{} consecutive images. Although the surface morphology might appear unchanged, a specific detail marked by a red dotted line reveals a transformation of the surface. It was noted that certain circular features had vanished, revealing previously covered circular features on the surface. To determine whether this characteristic is consistent throughout the scanning area, multiple scans were performed at the same spot before expanding the view to a larger area. In \textbf{Fig. \ref{fig:corrosion}b}, a distinct depression is visible within the black dashed lines, marking the area of the earlier scans. The line-trace shows that approximately \SI{1}{\nano\meter} of the film was removed by repeated scanning alone. We ruled out direct physical contact of the tip with the sample during scanning, as evidenced by the absence of current spikes and no alteration in lateral resolution attributable to changes in the tip's structure or radius. 

The correlation between the bias voltage and the evolution and degradation of surface roughness was investigated. The sequential imaging in \textbf{Fig. \ref{fig:corrosion}a} was performed at a low bias voltage of -\SI{25}{\milli\volt}. In \textbf{Figs. \ref{fig:corrosion}c - e}, the RMS roughness is given as a function of the frame number, depending on the tip bias. At a low tip bias of -\SI{25}{\milli\volt}, the first \SI{6} frames show a steady decrease in RMS roughness by a total of \SI{60}{\pico\meter}. It is probable that the highest circular features, which contribute to the increase in RMS roughness, are removed by the scanning. This process smooths the film and explains the reduction in RMS roughness. From frame number \SI{7}{} onward, the RMS roughness evolution stabilizes at \SI{265}{\pico\meter}, a state we call stabilization. The tip was then moved to a new surface area and the bias voltage was raised to -\SI{200}{\milli\volt}. Consecutive imaging was again performed, and the consistent corrosion in RMS roughness continues through \SI{20} successive frames, as depicted in \textbf{Fig. \ref{fig:corrosion}d}. This indicates that at a tip bias of -\SI{200}{\milli\volt}, surface structural corrosion by the STM tip occurs more slowly, and the smoothing process is more gradual. The tip was relocated once more to a different area and the bias voltage increased to -\SI{350}{\milli\volt}. In \textbf{Fig. \ref{fig:corrosion}e}, it is observed that the degradation in roughness is minimal and it maintains its initial value of \SI{310}{\pico\meter}.

\subsubsection{Tip-induced LDOS modification}

\begin{figure}
    \centering
        \includegraphics[scale=0.5]{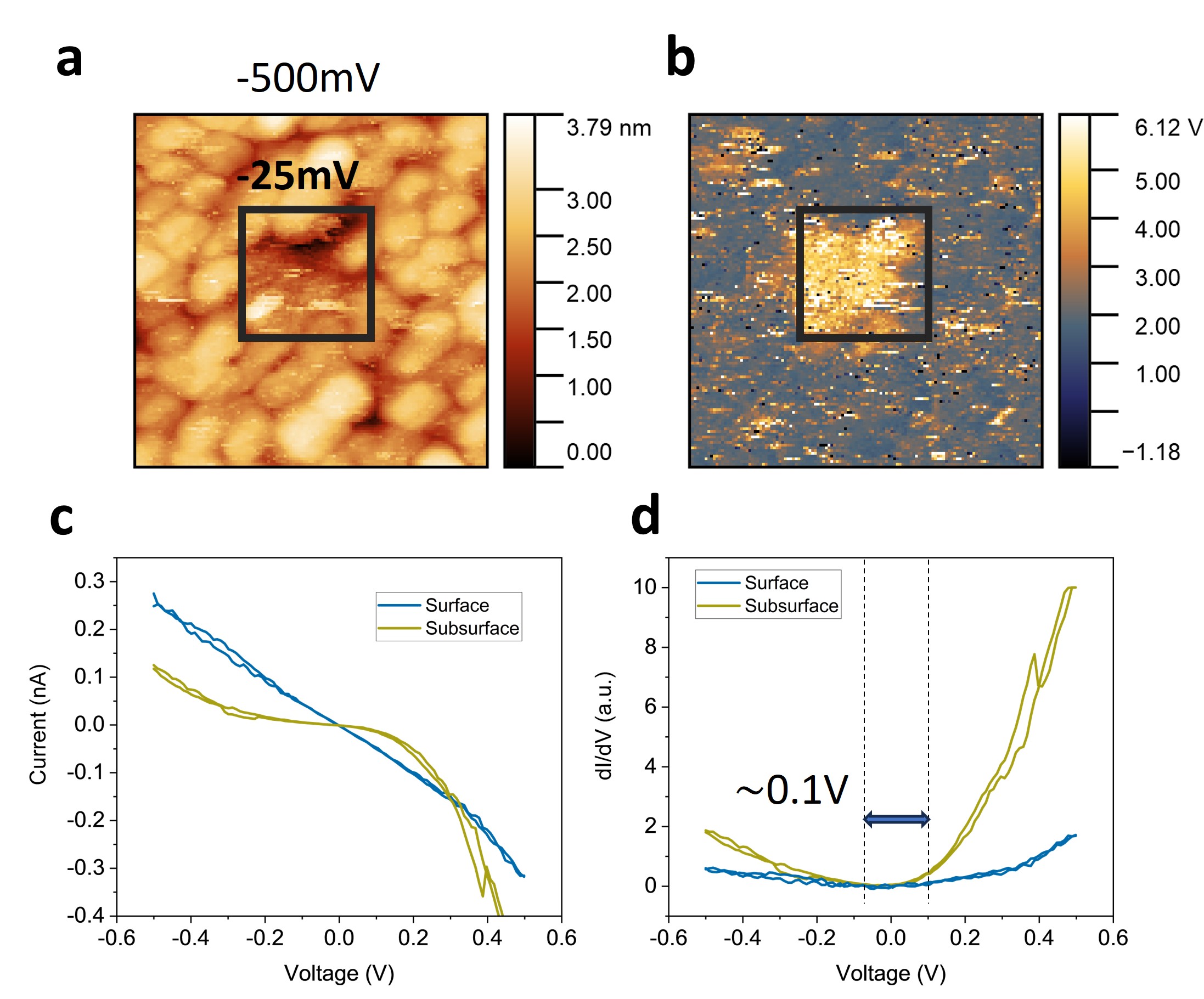}
            \caption{\textbf{Tip-induced STS change mapping} (\textbf{a}) Topographic STM image, tip bias -\SI{500}{\milli\volt}, tunnel current \SI{250}{\pico\ampere}. The black square indicates a \SI{25}{}x\SI{25}{\nano\meter} image taken at tip bias -\SI{25}{\milli\volt}, tunnel current \SI{250}{\pico\ampere}. (\textbf{b}) The tip-induced corrosion has been accompanied by a change in LDOS evident from the STS mapping, within the smaller scan area indicated with the black square. (\textbf{c, d}) $I(V)$ and d$I$/d$V$ STS taken at the pristine surface (blue) and the exposed subsurface (yellow) show a considerable change in LDOS from metallic to semiconductor with the formation of a $\approx$\SI{150}{\milli\volt} electronic gap at room temperature. The black scale bar is equal to \SI{10}{\nano\meter}.}
            \label{fig:corrosion_LNO_STS}
\end{figure}

We investigated the impact of LNO surface tip-induced degradation on the density of states (LDOS) using spectroscopic mapping. In \textbf{Fig. \ref{fig:corrosion_LNO_STS}a}, a topographic scan was conducted at a bias of -\SI{500}{\milli\volt}, following a detailed scan of a \SI{25}{\nano\meter} x \SI{25}{\nano\meter} region at a lower bias of -\SI{25}{\milli\volt}. This lower bias imaging caused local surface damage and exposed subsurface layers. Concurrently, we recorded the STS map at -\SI{500}{\milli\volt} as shown in Figure \ref{fig:corrosion_LNO_STS}b, which displayed a significant alteration in the LDOS between the subsurface and surface. The surface exhibits metallic characteristics, as confirmed by the $I(V)$ and d$I$/d$V$ STS plots in \textbf{Fig. \ref{fig:corrosion_LNO_STS}c and d}, which show no significant electronic gap. The exposed subsurface, due to tip-induced degradation, shows a semiconductor-like LDOS with the formation of a small gap of approximately \SI{100}{\milli\volt}. These findings demonstrate that tip-induced degradation has caused an irreversible alteration in the LDOS toward a semiconductor, from a metallic surface phase. 

In the study by Kotiuga \textit{et al.} \cite{Kotiuga2019CarrierVacancies}, oxygen vacancies were found to result in electron localization around the defect, causing a change in the valence state of Ni from Ni$^{3+}$ to Ni$^{2+}$, and significantly increasing resistivity. This observation is consistent with our local-formed electronic gap.

To elucidate the origin of tip-induced degradation, the role of oxygen vacancies is crucial. Near the surface, an elevated concentration of oxygen vacancies is expected \cite{Anjeline2022ProbingProperties, Arandiyan2021ImpactReaction}, which may be exacerbated by exposure to vacuum within the STM. The vacancy concentration is predicted to gradually decrease in depth in the film toward stoichiometry, as supported by the self-limiting surface degradation of \SI{1}{\nano\meter}. In tensile-strained LNO films, there is a preferential formation of oxygen vacancies at the apical oxygen sites \cite{Geisler2022OxygenSuperlattices}. These vacancies can locally alter the crystal lattice strain \cite{Zuev2008DefectLaCoO3d, Chen2005Thermal}. Possibly, with too many vacancies forming, the crystal can collapse. The loss of oxygen in the vacuum chamber may be transient and, hence, we cannot fully rule out the possibility that the surface oxygen stoichiometry has changed over the measurement period. This could also mean that the reproducibility of surface studies of LNO in a vacuum environment may be problematic because of the subtle dependence on the oxygen vacancy concentration. However, after 6 months in vacuum we still observe the same surface degradation effect. The irreversible effect of tip-induced oxygen extraction is probably dominant in a vacuum environment, where oxygen can no longer be replenished in the film.

The alteration in Ni-oxidation state can account for the transition in conductivity from metal to insulator upon introducing vacancies. Nevertheless, the study by Liao \textit{et al.} in Ref. \cite{Liao2021Oxygen3} has demonstrated that changes in Ni-oxidation state driven solely by oxygen vacancies are insufficient to describe the variations in resistivity. Instead, the emergence of a new phase characterized by alternating Ni-octahedra and planar-Ni in the oxygen-deficient phase is sufficient, leading to the development of a site-selective Mott-insulator and band-insulator, as supported by our STS measurements showing the formation of an electronic gap.

\subsubsection{Electric field gradient drives oxygen extraction and increased oxygen vacancy formation}

The formation of oxygen vacancies may be related to the electric field gradient between the tip of the STM and the sample surface \cite{Kalinin2011RoleFilms}. The tunneling current itself is unlikely to introduce surface degradation, as in our consecutive imaging experiments we always kept the current constant. The electric field between the tip and the sample is on the order of \SI{e8}{\volt\per\meter} and this could lead to the leeching of oxygen from the crystal structure. Evidence of electric field-driven migration of oxygen vacancies has been found in manganites \cite{Liao2012EvidencePr0.7Ca0.3MnO3}, for example. It should be noted that surface degradation is a feature limited to a thin layer on the surface. This could imply that the oxygen vacancy concentration is not uniform in the thickness of the film. 

From these results, it becomes evident that increasing the bias voltage can decrease the degradation of the surface. With STM, raising the bias voltage on an STM tip while maintaining a constant tunneling current results in an adjustment in the distance between the tip and the sample. Consequently, the electric field gradient between the tip and the sample varies, and this field (gradient) can exert a force on oxygen. We utilized a COMSOL Multiphysics FEM simulation to numerically calculate the electric field (gradient) between a tip and a sample. The model is given in \textbf{Supplementary Fig. S1}. The bias voltage varied between \SI{25}{\milli\volt}, \SI{200}{\milli\volt}, and \SI{300}{\milli\volt}. To maintain a constant tunnel current, the distance between the tip and the sample was varied to \SI{0.2}{\nano\meter}, \SI{1.5}{\nano\meter}, and \SI{2.0}{\nano\meter}, taken from the piezo responses observed in the STM. The results in \textbf{Supplementary Fig. S1} indicate that the vertical electric field, $E_z$, diminishes more strongly in the horizontal direction, $r$, at greater distances and bias voltages. The numerical derivative, $dE_z/dr$, shown in \textbf{Supplementary Fig. S1}, reveals that an enhanced electric field gradient is calculated for lower bias voltages. Therefore, a force (gradient) exerted on the vacancies leads to oxygen anion leeching from the surface into the vacuum chamber. Furthermore, the tip is likely very close to the surface when the electronic gap is formed by the further removal of oxygen vacancies, thereby an additional electrostatic force is exerted on the surface, which can lead to further surface damage. 

A rapid alteration in crystalline volume due to increased vacancies can lead to the disintegration of LNO circular features, visible via STM. In epitaxially strained films, the lattice expansion due to oxygen vacancies likely induced an expansion anisotropy \cite{Tyunina2021AnisotropicFilms}. In the LNO bulk, where oxygen vacancy levels are reduced, the film shows greater resistance to degradation induced by the STM tip.

\subsubsection{Evidence of oxygen vacancy clustering on the surface of LNO}

\begin{figure*}
    \centering
        \includegraphics[scale=0.35]{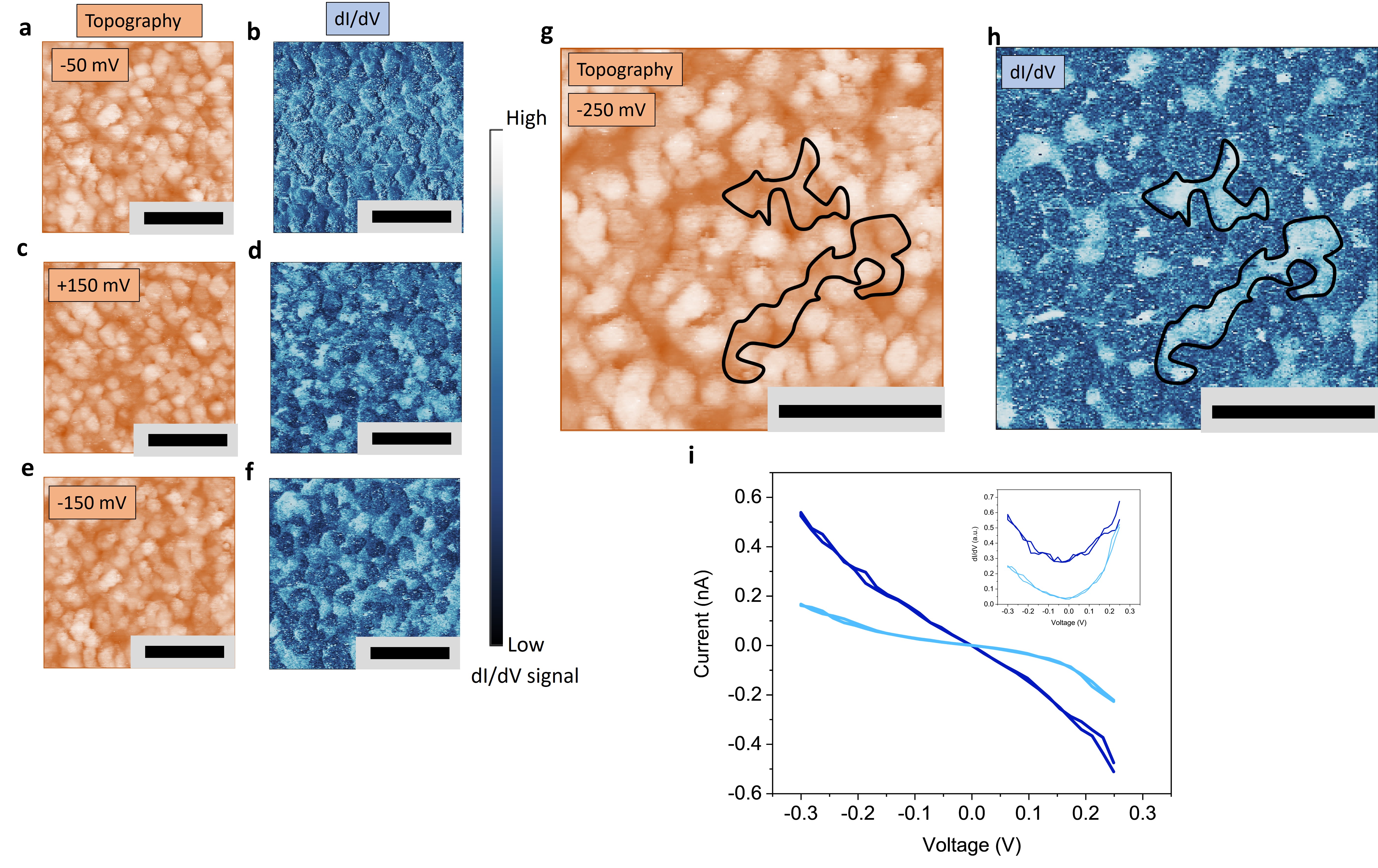}
            \caption{\textbf{STM and STS of LNO surface at different bias voltage}. (\textbf{a}, \textbf{c}, \textbf{e} and \textbf{g}) STM topography with variable bias voltage with a tunnel current of \SI{250}{\pico\ampere}. (\textbf{b}, \textbf{d}, \textbf{f} and \textbf{h}) Corresponding d$I$/d$V$ STS map as a function of bias voltage. The black scale bar is equal to \SI{40}{\nano\meter}. (\textbf{i}) $I(V)$ and d$I$/d$V$ STS spectra (inset) showing a metal and semi-conductive electronic phase taken on the blue and light blue regions of \textbf{h}, respectively.}
            \label{fig:STS}
\end{figure*}

In complex oxide perovskites, theoretical studies increasingly suggest that oxygen vacancies tend to cluster, forming distinct ordered phases \cite{Cuong2007OxygenStudy, shin_magnetic_2022, Aschauer2013Straincontrolled3}.  In this work, our STM and imaging bias-dependent STS mapping revealed the presence of oxygen vacancy clustering on the surface.  We present the results of this mapping, performed at various bias voltages, in \textbf{Figs. \ref{fig:STS}b, d, f, and h}, with the corresponding topographies shown in \textbf{Figs. \ref{fig:STS}a, c, e, and g}. At a bias voltage of $-$\SI{50}{\milli\volt}, a topography-dependent d$I$/d$V$ STS contrast was observed, arising from feedback-induced convolution near the circular feature boundary. Near a circular feature boundary, the need for the tip to quickly adjust its height, combined with the transient limitations of the feedback loop's speed, leads to a slight change in current, which is also detected by the lock-in during the STS mapping. This effect was minimized by scanning at slower speeds and optimizing feedback loop gains, with further cross-talk minimization given in \textbf{Supplementary S2}. 

Upon increasing the bias voltage to $\pm$\SI{150}{\milli\volt} and $-$\SI{250}{\milli\volt}, a distinct two-state LDOS emerged, resembling electronic phase separation, as illustrated in \textbf{Fig. \ref{fig:STS}f, h}. This phenomenon was characterized by regions displaying either elevated or reduced LDOS. These patches of LDOS appeared to have a slight connection with the local circular feature topography, although no clear relationship was found between the topographic features and LDOS intensity. It is established that an increase in oxygen vacancy concentration $\delta$ increases electrical resistivity \cite{Gayathri1998ElectronicStoichiometry, Kotiuga2019CarrierVacancies}. STS performed on the islands (light blue) and matrix (dark blue) are presented in \textbf{Fig. \ref{fig:STS}j}. The islands exhibited reduced tunneling LDOS and the emergence of a slight electronic gap. The adjacent matrix showed metallic properties, corroborated by the linear $I(V)$ trends observed in tunneling spectroscopy. The two-phase electronic structure can be indicated by a bimodal distribution, \textbf{Supplementary S3}. The distribution of oxygen vacancies by clustering on the surface can account for the observed islands with a small electronic gap. 

The appearance of LDOS irregularities around $\pm$\SI{150}{\milli\volt} is captivating, particularly in connection with the ARPES studies by Yoo \textit{et al.} \cite{Yoo2015LatentTopology} suggest the presence of a charge density (CD) instability at the nanoscale. The analysis of band structures, both theoretical and experimental, for tensile-strained LNO films reveals a flat band near $-$\SI{150}{\milli\volt} relative to the Fermi level. This flat band \cite{Malashevich2015FirstprinciplesStructures} suggests significant electron-electron interactions. Such instabilities are linked to the formation of CD and even spin-density waves in tensile-strained LNO films by Yoo \textit{et al.} \cite{Yoo2015LatentTopology}. Therefore, the relative change in local LDOS from a metal phase to a gapped phase in our STS works likely corroborates the existence of an electron-correlated phase above \SI{150}{\milli\volt}. In-depth investigations of strong electron correlation effects in LNO were conducted using DMFT and ARPES by Deng \textit{et al.} in Ref. \cite{Deng2012HallmarkBands}, revealing a distinctive kink in the band dispersion at -\SI{0.2}{\electronvolt}. Euguchi \textit{et al.} \cite{Eguchi2009Fermi3} also observed a similar kink at \SI{250}{\milli\electronvolt} in their ARPES measurements, which they ascribe to momentum-dependent mass renormalization and asymmetry between electron and hole in a Fermi liquid. Horiba \textit{et al.} \cite{Horiba2007ElectronicStudy} provide further evidence from soft x-ray photo-emission spectroscopy (PES) for the formation of correlated Ni 3d $e_g$ electrons near the Fermi-level.  Insights into electron interaction of LNO with other degrees of freedom, with these kinks occurring at up to several hundred millivolts \cite{Byczuk2007KinksElectrons}, have therefore been proficient in spectroscopy-based studies, and our work demonstrates the spatial aspects on the surface of LNO. Overall, these electron correlations result in deviations from the Fermi-liquid-like behavior in LNO. 

Furthermore, the PES spectra referenced in \cite{Horiba2007ElectronicStudy} were theoretically analyzed and well replicated by Liao \textit{et al.} \cite{Liao2021Oxygen3} employing DMFT. The ground state of LNO, determined by DMFT, was identified as metallic, which represents the surface electronic structure.  Introducing vacancies to create LaNiO$_{2.5}$, can lead to an alternating structure of Ni-O octahedra and Ni-O square planes \cite{Liao2021Oxygen3}, missing the apical oxygens. These ordered vacancy structures have also been observed by HRTEM \cite{Lopez-Conesa2017EvidenceFilms} in bulk LNO thin films. The DMFT DOS of Ref. \cite{Liao2021Oxygen3} shows that a localized and sharp e$_g$ peak exists below \SI{0.5}{\electronvolt}. This is supported by Sawatzky $\&$ Green \cite{Pavarini2016Quantum2016} and Li \textit{et al.} \cite{Li2021SuddenFilms} and assigned to a 3$d^8\underline{L}$ state, due to strong electron affinity of the late 3d elements like Ni in combination with the sharing of the oxygen vacancies between the octahedral and plane NiO unit, thereby increasing the 3d-O2p holes hybridization. As discussed and supported by \textbf{Fig. \ref{fig:STS}}, the islands on the surface is on a scale of \SI{20}{}-\SI{40}{\nano\meter}, representing individual regions, associated with this electronic phase separation. The STS results show the distinct difference for the polarity of the bias voltage, probing the filled e$_g$ states close to the Fermi-level and the O-2p broadened empty states respectively. In the production of OER process, the O2 takes place by O- formation on the surface, supplying an electron to the empty mainly O-2p states. This might weaken the Ni3d-O2p bond and creates an O vacancy.

In our work, we observe that the STS mapping represents high/low LDOS by dark  blue/light blue regions in \textbf{Fig. \ref{fig:STS}d} and \textbf{\ref{fig:STS}f} for the filled and empty LDOS, respectively. The dark blue regions show for both (filled and empty) low binding energy states and around $E_F$, corresponding with the “ordered” octahedral NiO sites, supported by the more filled e$_gd_{z^2}$ and more empty e$_g$$d_{x^2 - y^2}$ states respectively, as in case of LNO$_{2.5}$, see \cite{Horiba2007ElectronicStudy}.

We performed XPS to further identify the effect of oxygen vacancies. The XPS-valance band are given in \textbf{Supplementary S4}. We did not observe a significant sharp peak a few \SI{100}{\milli\electronvolt} above E$_f$ as noted by Horiba \textit{et al.}\cite{Horiba2007ElectronicStudy}, who assign this to the e$_g$ of stochiometric LNO. Inducing vacancies in the film, by heating, the photoemission spectra in Ref. \cite{Horiba2007ElectronicStudy} lead to a decrease in amplitude and broadening of the e$_g$ peak near E$_f$. A complete disappearance the e$_g$ is observed for elevated heating temperatures with increased oxygen vacancy concentration. The authors also note that from the DFT calculations, the strain, tetragonal (P4\textit{mm}) for LNO grown on STO, can lead to the disappearance of the sharp e$_g$ peak. Therefore, our XPS data can be interpreted as a strained film (surface) with oxygen vacancies present, leading to a decrease in e$_g$ intensity. We do not observe a notable increase in the XPS valence band gap because of the presence of a metallic matrix on the surface that is mixed with the semiconductive islands, as observed with our STM. Furthermore, we cannot rule out the sensitivity limitations of our XPS compared to a beamline \cite{Horiba2007ElectronicStudy}.  

\section{Discussion}

\subsection{Effect of oxygen vacancies on the surface transformation of LNO pre-catalyst}

The tip-induced degradation effect can be associated with recent reports in surface transformations \cite{Baeumer2021TuningElectrolysis} and previous studies on hydrooxidation \cite{Mickevicius2009SurfaceFilms}. Surfaces prone to reconstruction, such as LNO, can facilitate the formation of nickel hydroxides, which have been reported to exhibit greater activity in the OER \cite{Baeumer2021TuningElectrolysis}. Consequently, oxygen vacancies might not directly enhance OER activity, but could enhance surface reconstruction and OH incorporation \cite{Zhao2020BoostingDecoration}. In this work, sharp morphological features are suggested to amplify electric field gradient and enhance oxygen vacancy formation rate. The electric field can furthermore promote OH-ion migration and aggregation, thereby accelerating OER kinetics \cite{Liu2021TipReactions}. Surface amorphization has been observed at defects such as Ruddlesden–Popper phases and pit-like structures \cite{Bak2017FormationNickelate}, where we propose that electric field gradients can be substantial enough to create significant oxygen vacancies, resulting in anisotropic crystal strain and surface transformations at high over-potential. Similar degradation driven by point defects has been noted for manganites \cite{Mierwaldt2017EnvironmentalCatalysts}. This implies that the surfaces of these films are intrinsically unstable \cite{Binninger2015ThermodynamicCatalysts}, particularly for surfaces with notable corrugation, like the circular features observed in LNO - which we consider to be a typical feature in many reported epitaxial nickelate films. Perfectly epitaxial films with atomically smooth surfaces, exhibiting corrugations at the atomic scale across the entire width of the plateaus, might experience a reduced rate of oxygen vacancy formation during OER.

An enhancement in OER activity has been observed for LNO catalysts as function of oxygen vacancies, yet gives also a increase in resistivity  \cite{McBean2017GeneralizableMedia}. Thus, an alternative mechanism must be responsible for the increased OER activity. The involvement of lattice oxygen in the lattice oxygen evolution mechanism (LOM) is known to be crucial for enhancing the OER \cite{Choi2023EpitaxialReaction}. Additionally, the formation of metal (oxy)hydroxide species on the surface can provide a stable and efficient catalyst \cite{Baeumer2021TuningElectrolysis}. Therefore, the as-grown surface of the LNO should be considered a pre-catalyst, with the actual catalyst forming during operando conditions. When exposed to the electrolyte and local electric field gradients near the circular feature boundaries, the oxygen vacancies can cause an unstable crystal structure, particularly near the surface. charge trapping near defects probably results in further lattice distoriton, which lead to the eventual collapse of the crystal \cite{Tyunina2021AnisotropicFilms}. Local OH$^-$ ions probably screen and reduce the electric field, potentially reducing the rate at which the electric field gradients can lead to the extraction of oxygen anions. The different rate between oxygen replenishment and leaching and formation of too many oxygen vacancies during OER likely leads to non-reversible surface transformation \cite{Pan2020DirectParticipation}. 
\section{Conclusions}

The findings of this study emphasize that PLD-grown epitaxial thin-film LaNiO$_{3-x}$ is a fascinating complex oxide perovskite, particularly in terms of its corrugated surface morphology, aggregation of oxygen vacancies, and emergent semiconducting clusters within a metallic matrix. LNO's outstanding catalytic performance in the oxygen evolution reaction calls for further investigation into the relationship between oxygen non-stoichiometry and the origin of in-operando surface transformations. We observed a nanoscale surface morphology consisting of coalescent, crystalline circular features. The identification of a two-phase LDOS is likely due to oxygen vacancy clustering. The electron correlation phenomena in LNO are known to show a kink and flatband in the band-dispersion between \SI{100} and \SI{250}{\milli\electronvolt}, which we observe and confirm with STS in real-space on the surface. Using an STM tip proves to be an effective method for inducing and examining the oxygen vacancies on the surface of nickelate films, which can be challenging to analyze using XPS or transport measurement techniques, relevant for electrochemistry and device applications. 

\subsection*{Author Affiliations}
Eindhoven University of Technology, The Netherlands. Department of Applied Physics and Science Education. Molecular Materials and Nanosystems (M2N), Spectrum, de Zaale \SI{5612}{} AP, Eindhoven. 

\section*{Methods}
The sample is grown with similar conditions as in Ref. \cite{Baeumer2021TuningElectrolysis}. The sample was grown at \SI{650}{\degree}C, making the surface lanthanum rich \cite{Baeumer2021TuningElectrolysis}. High quality epitaxial growth in the (001) direction was confirmed with XRD given in \textbf{Supplementary figure S5}.  

\subsection*{Ambient AFM}
Tapping-mode AFM was performed using a Nanosensor NCH-PPP Si cantilever on a Veeco Dimension III microscope under ambient conditions. Imaging was carried out with a scan speed of \SI{1} lines per second (512 lines/512 pixels). Post-processing of the AFM data was performed with Gwyddion software \cite{Necas2012GwyddionAnalysis}. Images were line aligned using median of differences and plane leveled using mean subtraction. 

\subsection*{UHV STM and STS}
Scanning tunneling microscopy was performed with a Scienta Omicron GmbH VT-SPM operating in an ultra high vacuum with a base pressure of \SI{e-10}{\milli\bar} at \SI{300}{\kelvin}.  STM tips were mechanically cut from PtIr wire. The high-resolution imaging capability of the tip was confirmed by observing the atomic resolution of SiC-graphene. The bias was applied to the tip and the sample was grounded. The imaging was performed in constant current mode. For d$I$/d$V$ (STS) mapping and spectroscopy, a Stanford Research Instruments SR830 lock-in amplifier with an alternating voltage between \SI{20}{\milli\volt} and \SI{100}{\milli\volt} applied at frequency above \SI{3}{\kilo\hertz}, within the long-pass filter of the phase lock loop. The crosstalk by capacitance was minimized prior to STS measurements. Lateral drift compensation was calibrated for consecutive imaging by measuring the $x$-$y$ drift between two images with the same topographic features. The drift compensation was fine-tuned during the measurement period as a result of small thermal fluctuations and stabilization of the piezotube drift. The post-processing of the STM data was performed with Gwyddion software \cite{Necas2012GwyddionAnalysis}. The images were line-aligned using median of differences and plane-levelled using mean subtraction. 

\subsection*{COMSOL Multiphysics FEM simulation}
COMSOL Multiphysics was used. A tip was modeled with a radius of \SI{5}{\nano\meter}. The voltage was applied to the tip and the resulting electric field was calculated using the AC-DC package. The mesh grid was incrementally smaller toward the tip apex to account for nanometer-scale tip-sample separation by providing enough mesh resolution. The sample was held at ground potential.

\subsection*{XPS}
XPS spectra were acquired using a Thermo Scientific K-Alpha XPS system. The source emitted Al K$\alpha$ radiation with an energy of 1486.6 eV, and a monochromator was used to selectively transmit these x-rays. The XPS instrument has a work function of \SI{4.5}{\electronvolt}. To ensure a sufficient signal-to-noise ratio, 30 scans were recorded and averaged for the oxygen 1s spectra, while 50 scans were taken for the metal and valence spectra. The dwell time per point was set to \SI{50}{\milli\second}. Spectra were processed and analyzed using CasaXPS software. Shirley background subtraction was applied and the peaks were fitted with a combination of 70\% Gaussian and 30\% Lorentzian profiles.

\section*{Acknowledgement}
We would like to acknowledge financial support by the Eindhoven University of Technology. Furthermore, we thank dr. Christoph Beaumer and Emma van der Minne of the University of Twente, Netherlands, for growing the LNO sample. We thank Laura Donk of the Eindhoven University of Technology, department of Chemical Engineering and Chemistry in the group Inorganic Materials and Catalysis for the XPS assistance. We thank Ries Koolen of the Eindhoven University of Technology, department of Applied Physics and Science Education for the XRD measurement.

\bibliography{references.bib}

\end{document}


\subsection*{S1 - Electric field gradient modeling}

\begin{figure}
    \centering
        \includegraphics[scale=0.4]{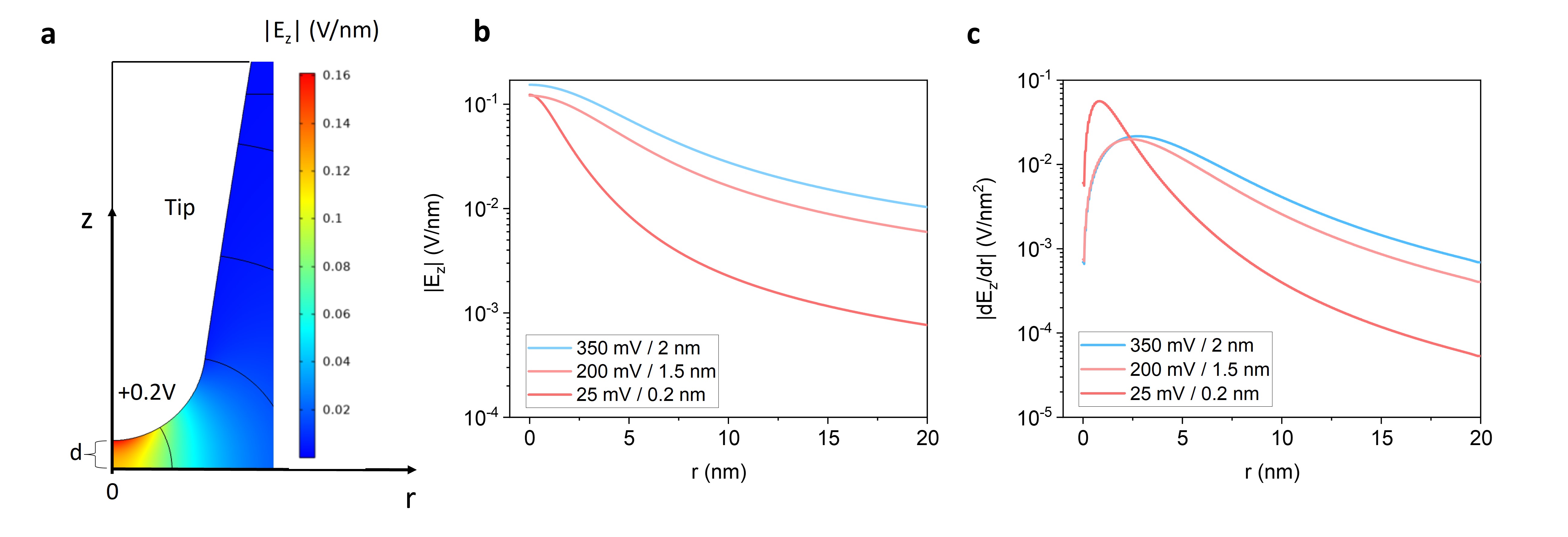}
            \caption{\textbf{COMSOL Multiphysics numerical model of the electric field between tip and sample.} (\textbf{a}) Electric field in $z$ direction ($E_z$) below the tip at a height $d$. The bias is applied to the tip with the surface grounded. (\textbf{b}) Calculated $E_z$ as a function of $r$ with different bias voltage and the corresponding tip-sample distance $d$ (nm). (\textbf{c}) Numerically calculated electric field gradient $dE_z/dr$ of \textbf{b}.}
            \label{fig:Electric field}
\end{figure}

A Finite Element Model using COMSOL Multiphysics was developed to compute the electric field and its gradient between an STM tip and a sample. The tip is represented as a half-structure due to its rotational symmetry and is projected in cylindrical coordinates, as shown in \textbf{Fig. \ref{fig:Electric field}a}. The distance $d$ between the tip and the sample is specified and ranges from \SI{2}{\nano\meter} to \SI{0.2}{\nano\meter}. In \textbf{Fig. \ref{fig:Electric field}a}, the computed electric field $E_z$ is shown. The flat surface is located at $z=0$, and the sample extends radially outward with $r$. A bias voltage of \SI{0.2}{\volt} is applied to the tip, while the sample is grounded. The vertical electric field $E_z$ as a function of $r$ on the surface of the sample is shown in \textbf{Fig. \ref{fig:Electric field}b}. The numerically computed field gradient that extends radially outward, $dE_z/dr$, is presented in \textbf{Fig. \ref{fig:Electric field}c}.

\subsection*{S2 - Excluding STS artifacts}

\begin{figure}
    \centering
        \includegraphics[width=\columnwidth]{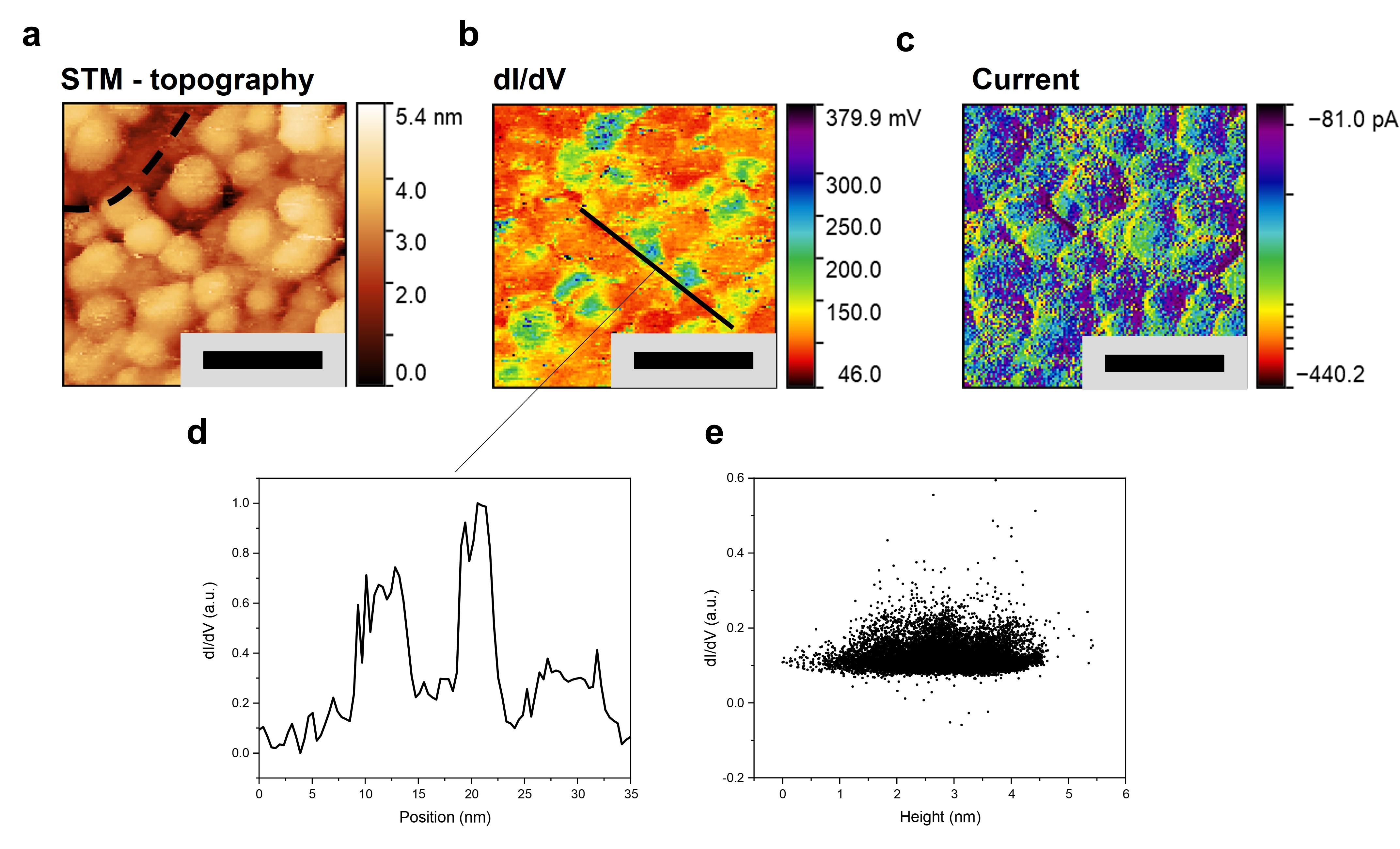}
            \caption{\textbf{Excluding d$I$/d$V$ artifacts.} (\textbf{a}) Topographic image, \SI{250}{\milli\volt}, \SI{150}{\pico\ampere}. A step edge is indicated with the black dashed line. (\textbf{b}) Corresponding d$I$/d$V$ map. (\textbf{c}) Corresponding current map taken with the feedback at constant current mode.  The scale bar is equal to \SI{5}{\nano\meter}. (\textbf{d}) dI/dV line-trace of \textbf{b}. (\textbf{e}) correlation between d$I$/d$V$ and height of \textbf{b}. }
            \label{fig:STS_Comparison}
\end{figure}

Topographic and/or electronic crosstalk could be excluded as the origin of the LDOS variation as a function of the oxygen vacancy clustering variation across the surface of the LNO thin film. In \textbf{Fig. \ref{fig:STS_Comparison}a} a STM \SI{45}{}$\times$\SI{45}{\nano\meter} topographic image (\SI{250}{\milli\volt}, \SI{150}{\pico\ampere}) is given close to a step edge, as indicated by the dashed black line. The grains are clearly resolved. We performed simultaneous d$I$/d$V$ STS mapping at the same bias voltage, \textbf{Fig. \ref{fig:STS_Comparison}b}, which shows a distinction in local DOS, with red regions having larger LDOS than green / blue regions. When examining the current map, \textbf{Fig. \ref{fig:STS_Comparison}c}, little to no correlation between d$I$/d$V$ STS and the current, indicating that there has been negligible crosstalk between topographic feedback and spectroscopy. The cross-section of \textbf{Fig. \ref{fig:STS_Comparison}d} shows that the variation of LDOS can be observed as a two-level signal. The correlation between LDOS and height is excluded to drive the contrast, \textbf{Fig. \ref{fig:STS_Comparison}e}.

\subsection*{S3 - Bimodal distribution of the vacancy clustering}

\begin{figure}
    \centering
        \includegraphics[width=\columnwidth]{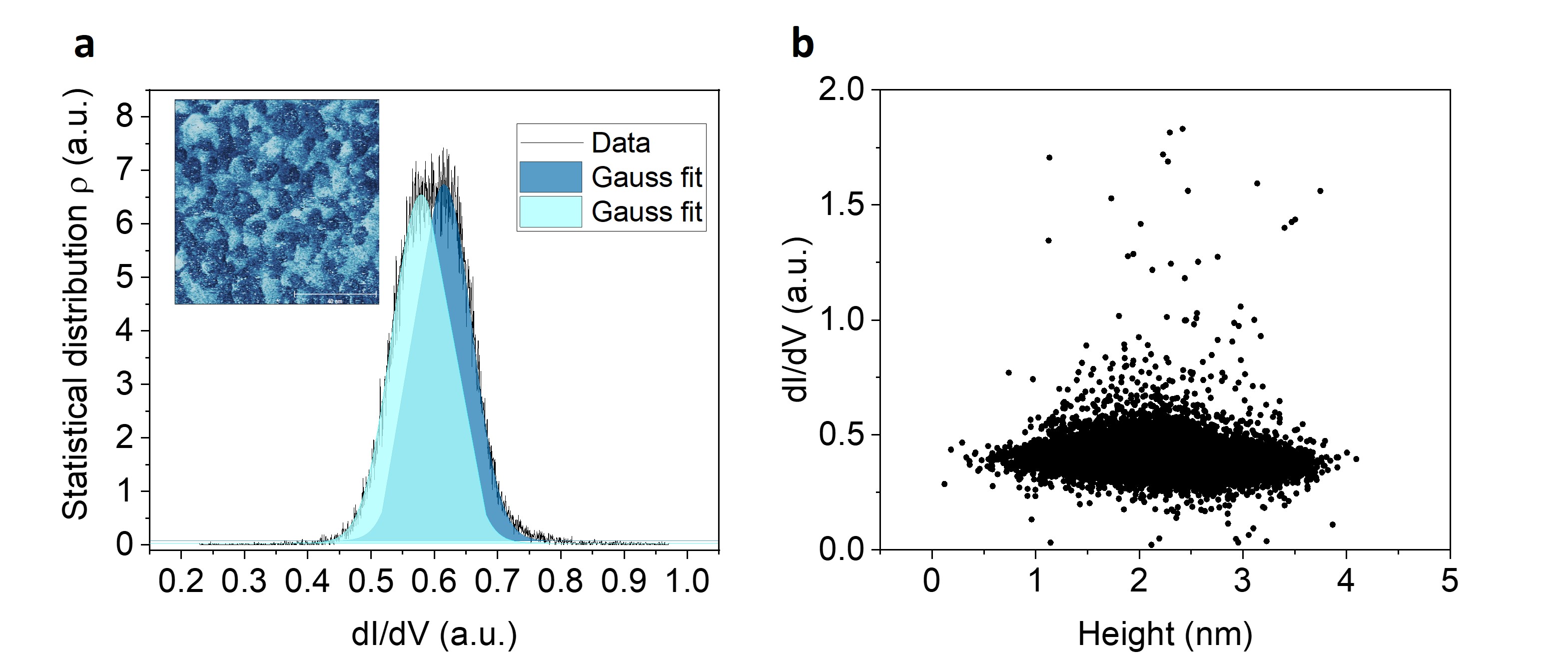}
            \caption{\textbf{Statistical analysis of the STS showing a bimodal distribution and no correlation between the local height and LDOS.} (\textbf{a}) d$I$/d$V$ STS statistical distribution ($\rho$) showing a bimodal distribution as fitted with two Gaussian curves. The inset shows the d$I$/d$V$ STS map used for the analysis. (\textbf{b}) Correlation between local height and d$I$/d$V$ STS signal of each image pixel.}
            \label{fig:STS_analysis}
\end{figure}

A two-phase surface electronic structure can be identified by a bimodal distribution, \textbf{\ref{fig:STS_analysis}}. The distribution of oxygen vacancies by clustering on the surface can account for the observed islands with a small electronic gap.

\subsection*{S4 - Valence band XPS}
\begin{figure}
    \centering
        \includegraphics[scale=0.5]{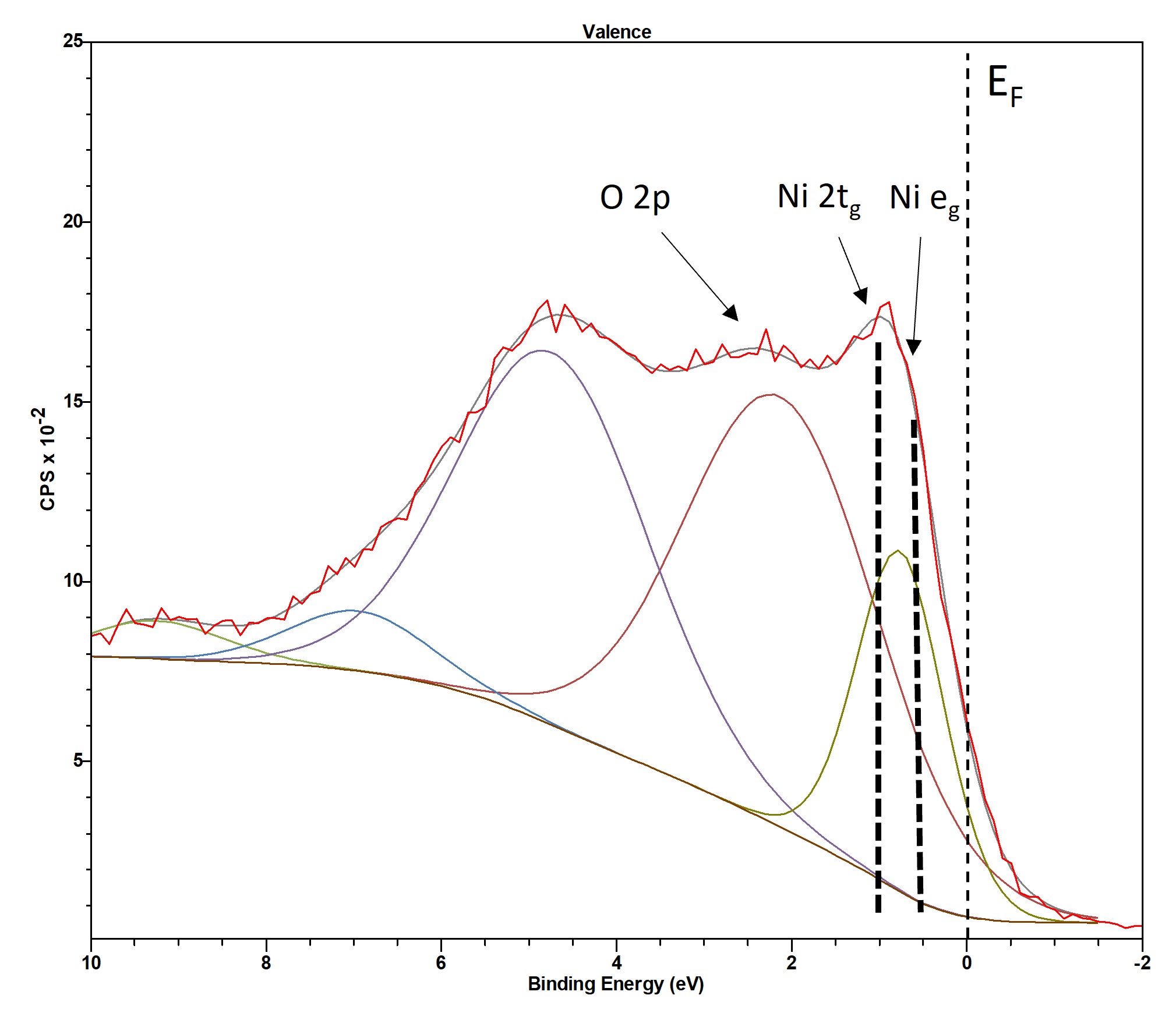}
            \caption{\textbf{Valence band XPS}.}
            \label{fig:LNO_Valence_XPS}
\end{figure}

Valence band spectra of the \SI{20}{\nano\meter} thick LNO film deposited on STO (001) substrate, given in \textbf{Fig. \ref{fig:LNO_Valence_XPS}}. 

\subsection*{S5 - XRD}
\begin{figure}
    \centering
        \includegraphics[scale=0.7]{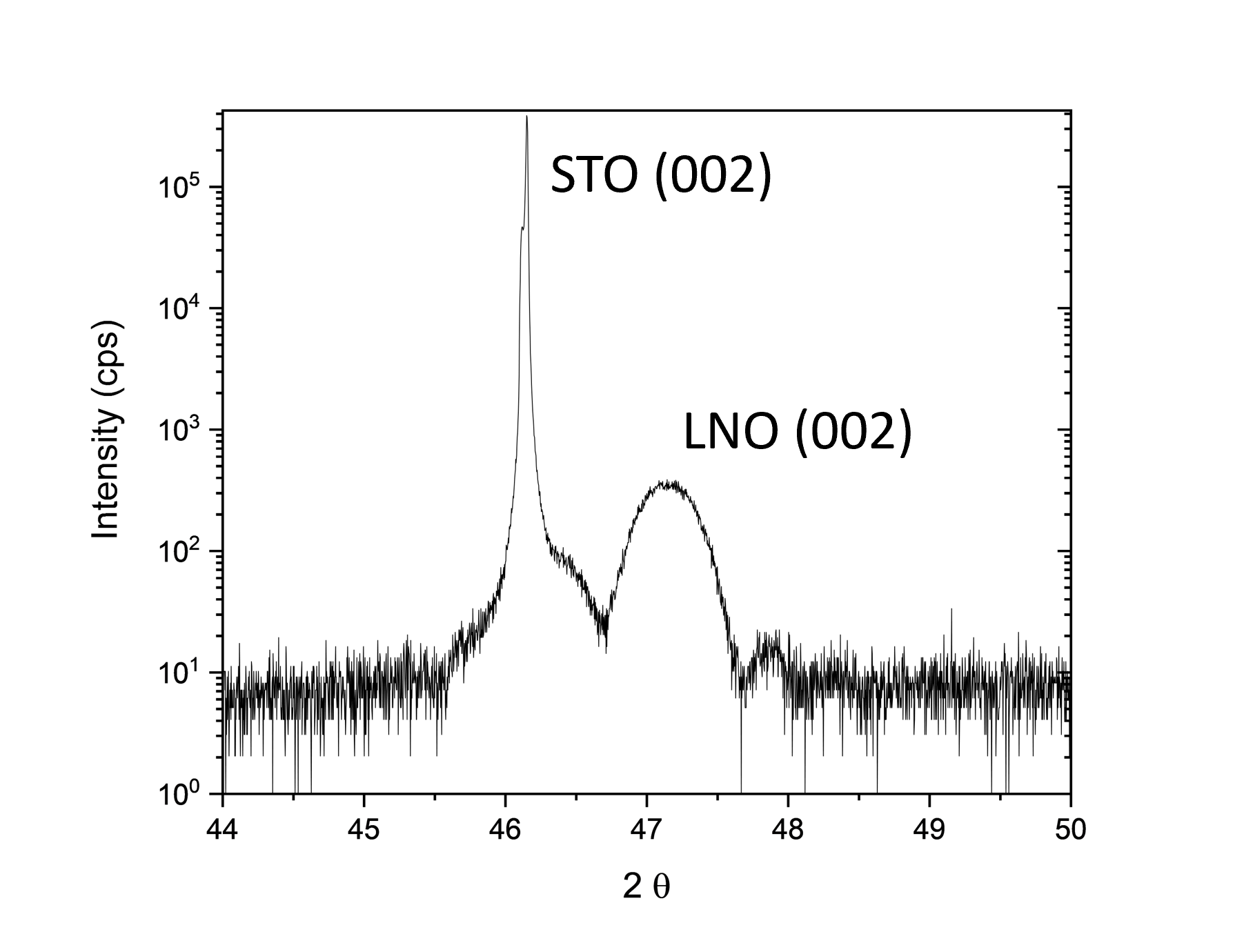}
            \caption{\textbf{X-ray diffractogram} Spectra around the (002) peak of the \SI{20}{\nano\meter} thick LNO film deposited on STO (001) substrate, indicating (001) orientation of the film.}
            \label{fig:XRD}
\end{figure}

XRD spectra around the (002) peak of the \SI{20}{\nano\meter} thick LNO film deposited on STO (001) substrate, indicating (001) orientation of the film, given in \textbf{Fig. \ref{fig:XRD}}. 

\newpage


\newpage

\newpage

\newpage

